Article:

by Donald Salisbury, Department of Physics, Austin College

# In Memoriam: Joshua N. Goldberg

Joshua N. Goldberg was one of the leading general relativists of the twentieth century. He made major contributions in the development of the theory describing the generation and propagation of gravitational waves, and in his capacity as an administrator of US Air Force funding he was largely responsible for the financial support of research in general relativity in the late 1950's and early 1960's, a period now recognized as a Renaissance era of general relativity. The Goldberg name had been assigned to his Russian emigrant father at Ellis Island. The original Jewish name was Sclaroff[1]. Josh was born in Chicago on May 30, 1925. He passed away on October 5, 2020.

I was privileged to have known Josh both as a teacher and a friend, beginning when I entered the graduate physics program at Syracuse University in the Fall of 1968. I will include here some of my personal reminiscences, supplemented with a series of interviews that I recorded with him over the years. A broader overview of his work can be found in an upcoming book[2]. I still recall the gracious invitation to visit his home that he and his wife Gloria made to us incoming graduate students that Fall. Josh had returned to Syracuse in 1963, having left after obtaining his PhD in relativity under the direction of Peter G. Bergmann in 1952. He was in fact one of Peter's first students. I was fortunate to have been introduced to relativity in a two semester graduate course taught by Josh in the academic year 1972-73. It played a major role in my decision to complete my thesis in this field under Bergmann's supervision.

The central focus of Josh's thesis was the explicit demonstration of the link between the general covariance of general relativity and the iterative approximation procedure that had been introduced by Einstein, Infeld, and Hoffmann to deduce particle equations of motion. The EIH procedure was based on what they called 'the lemma', a consequence of the fact that the field equations contained an anti-symmetric contribution. Bergmann made the first step in 1949 in tracing the origin of this contribution to general covariance. Josh exploited the fact that general covariance leads to a strongly conserved pseudo stress-energy tensor. Strong conservation means that its divergence vanishes whether the field equations are satisfied or not. It follows that this pseudo tensor can in turn be written as an antisymmetric superpotential, and this antisymmetry was precisely what he had sought to find.

In 1952 he went to work at the Armour Research Foundation in Chicago. He was able to find time while working on applied physics problems to continue with his thesis focus. Much came to fruition after securing his position at the Aerospace Research Laboratories (ARL) in 1954. Major innovations included a focus on the role of the Riemann tensor in identifying gravitational radiation and a correct tracking of coordinate conditions in *v/c* expansions in work on source equations of motion[3]. 1954 was the year in which he had the opportunity to discuss his work with Einstein. "And I explained to him the issue of strong conservation, which he probably knew, but had never expressed it. But nonetheless, I thought I was telling him something new. I've always afterward been a little embarrassed by that thought. But, so it goes. Anyway, he did seem interested. But he was not really interested in equations of motion. If I come to him and show him a plane wave solution, any wave solution, he would have been ecstatic, but he was not interested in radiation from moving particles."[4] At ARL he worked in 1962 with Peter Havas on a `fast motion` approximation, with a pioneering treatment of Green function divergences. This same year he obtained a leave to work as a National Science Foundation Fellow at the University of London. There he joined forces with another former Bergmann student, Ray Sachs, to achieve his likely best known theoretical result, the Goldberg-Sachs theorem[5]. Prior to 1962 it had been known that all algebraically special solutions of Einstein's vacuum equations possessed shear-free null congruences. The theorem proved the converse.

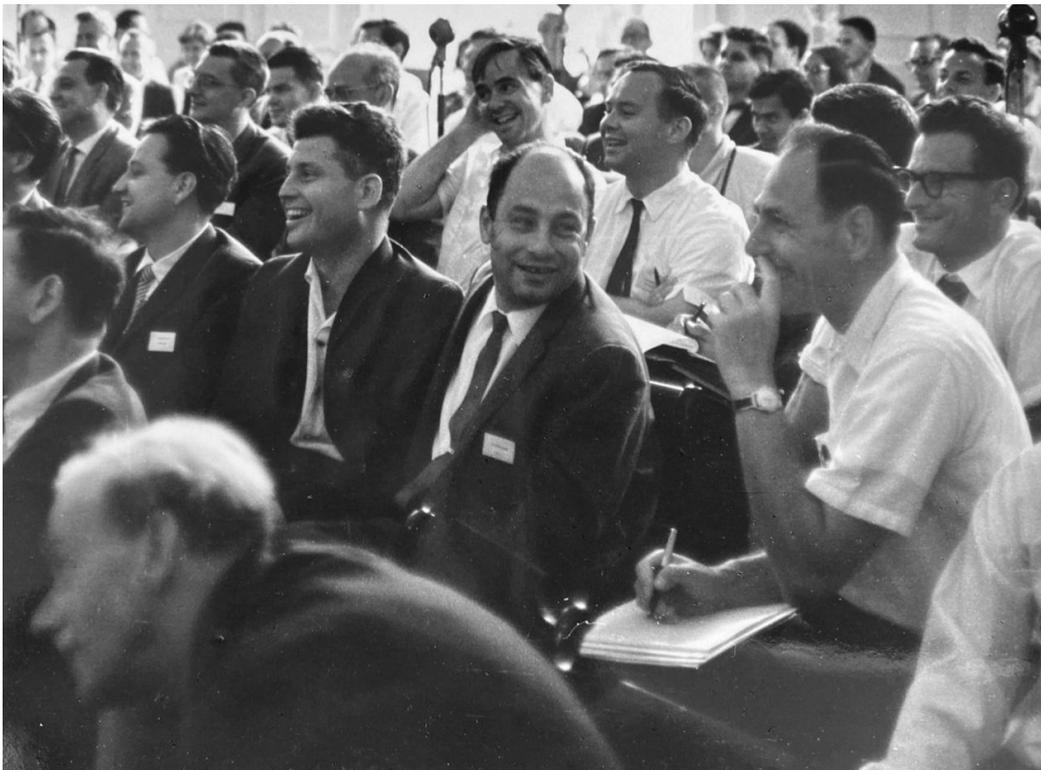

Art Komar, Peter Bergmann, and Joshua Goldberg at the 1962 Warsaw relativity meeting. Paul Dirac in the foreground. (Photo courtesy of Joanna Robinson)

New discoveries did of course not cease on his return to Syracuse. Disputes on the production and description of gravitational waves persisted. Null geodesics and null surfaces also remained a focus. Much was inspired by Sach's pioneering research in which Petrov types fall off along null rays with varying inverse powers of the radial coordinate. It was natural in this context to adapt the Newman-Penrose formalism to the investigation of the asymptotic behavior of fields, and this led to his most cited work[6]. In 1976 he co-authored an influential paper with Jürgen Ehlers, Havas and Arnold Rosenblum that summarized the inadequacies of all the current gravitational back reaction procedures[7]. Also, his familiarity with Ted Newman's asymptotic *H-* space and Penrose's twistor program led him to discover a new geometrical approach to Abhay Ashtekar's gravitational field variables. The outcome in 1988 was a new derivation of the Ashtekar constraint algebra and the associated symmetry transformations[8]. Abhay had first joined the relativity group at Syracuse in 1980, assuming a position as Assistant Professor that had been created with an NSF grant that Josh had marshaled through the system while serving as the Physics Department Chair.  It was natural, given the long-standing emphasis on the Syracuse group on the canonical quantization of Einstein's theory, that in the late 1980's Josh would turn his attention, with collaborators David Robinson and Chrys Soteriou, to the canonical formalism on null surfaces[9]. Finally, as Abhay has testified[2], Josh produced in 1992 a landmark paper with Jerzy Lewandowski and Cosimo Stornailo dealing with the self-dual gravitational connection that put the loop quantum gravity approach on a firm footing[10].

Josh's role as an administrator of Air Force funding in general relativity is well known and much appreciated. Less well known are his achievements as the Chair of the Syracuse Physics Department where with some internal opposition he managed the installation of new laboratories within the physics building. And I personally thank him for having extended my teaching assistantship appointment beyond the usual six year limit!

He was also, of course, an active presence in weekly seminars and regular work sessions with relativity graduate students, resident post-docs, and frequent visitors. I recall many instances in which he pursued calculations at the relativity group blackboard. His demeanor was always calm, confident, and poised. He radiated a non-pretentious self-confidence and never hesitated to engage if he disagreed. There was never any doubt that he was simply seeking the truth.

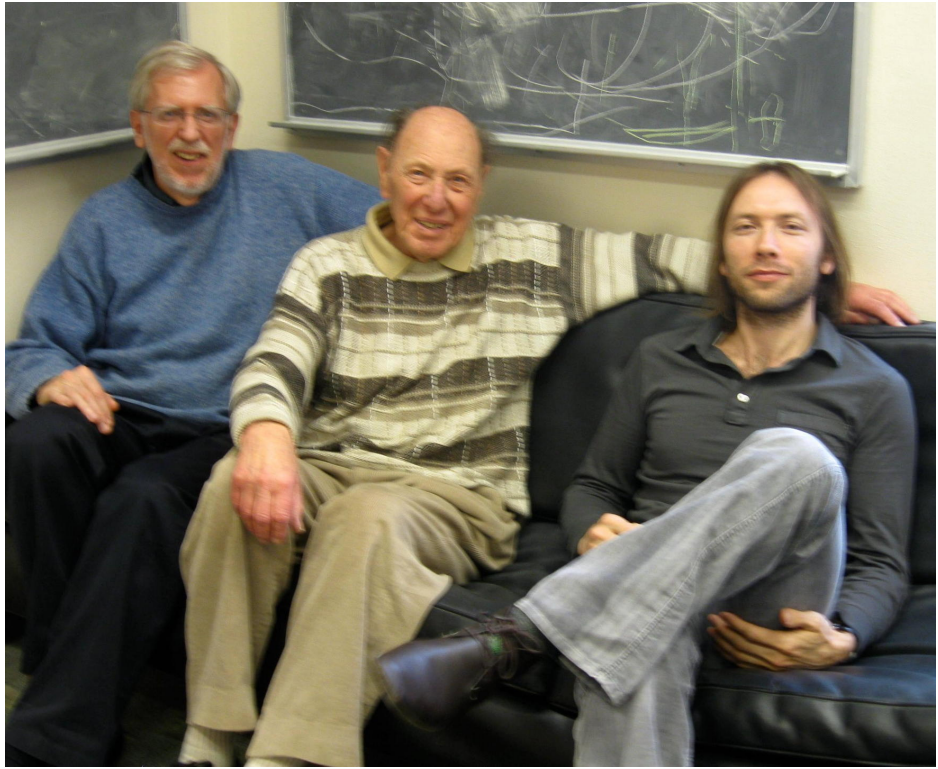

D. Salisbury, Joshua Goldberg, and D. Rickles – Syracuse, March, 2011

We did indeed disagree on an issue that had occupied the Syracuse relativity school since Bergmann's arrival in 1947[11]. Josh believed that an overarching principle was still lacking, and in particular, "anything we find may be in conflict with general covariance and I'm not, and I hope that my friends are not die-hards so that they'll insist that general covariance is an ultimate requirement that they refuse to give up in any case"[12]. History will be the ultimate judge!

**Notes and References**

[1] Oral History interview of Joshua Goldberg by Dean Rickles and Donald Salisbury, March 21, 2011, https://www.aip.org/history-programs/niels-bohr-library/oral-histories/34461

[2] D. Salisbury, A. Ashtekar, and D. Robinson, "Joshua N. Goldberg" in D. Malafarina and S. Scott, eds., *The Golden Age of Space-time,* (2021)

[3] J. N. Goldberg, "Gravitational radiation", Phys. Rev., **99**, 1873 (1955).

[4] Interview in Columbus, Ohio, April 2018

[5] J. N. Goldberg and R. K. Sachs. "A Theorem on Petrov types", Acta Physica Polonica, **22**, 13 (1962); reprinted in Gen. Rel. Grav., **41**, 433 (2009).

[6] J. N. Goldberg, A. J. Mcfarlane, E. T. Newman, F. Rohrlich, and E. C. G. Sudarshan, "Spins spherical harmomics and eth", J. Math. Phys., **8**, 2155 (1967).